\documentclass[twocolumn,showpacs]{revtex4}

\usepackage{graphicx}
\usepackage{dcolumn}
\usepackage{amsmath}

\usepackage[samesize]{cancel}

\begin{document}

\title{Zero-energy modes, fractional fermion numbers and the index theorem
in a vortex-Dirac fermion system
}

\author{Takashi Yanagisawa
}

\affiliation{
National Institute of Advanced Industrial Science and Technology
1-1-1 Umezono, Tsukuba, Ibaraki 305-8568, Japan; t-yanagisawa@aist.go.jp
}

\begin{abstract}
Physics of topological materials have attracted much attention from both
physicists and mathematicians recently.
The index and the fermion number of Dirac fermions play an important
role in topological insulators and topological superconductors.
A zero-energy mode exists when Dirac fermions couple to objects with soliton-like
structure such as kinks, vortices, monopoles, strings and branes. 
We discuss a system of Dirac fermions interacting with a vortex and
a kink.  This kind of systems will be realized on the surface of
topological insulators where Dirac fermions exist.
The fermion number is fractionalized and this is related to the
presence of fermion zero-energy excitation modes.
A zero-energy mode can be regarded as a Majorana fermion mode when the
chemical potential vanishes.
Our discussion includes the case where there is a half-flux quantum vortex
associated with a kink in a magnetic field in a bilayer superconductor.
A normalizable wave function of fermion zero-energy mode does not exist
in the core of the half-flux quantum vortex.
The index of Dirac operator and the fermion number have additional contributions
when a soliton scalar field has a singularity.
\end{abstract}


\maketitle

\section{Introduction}

Recently, topological materials have been attracted much attention
in physics.
New interesting topological properties will emerge in the study of
quantum systems from the viewpoint of topology.
In topological materials, Dirac fermions sometimes exist on the surface or
in the bulk.
The index of Dirac operators plays an important role in the study of 
topological systems\cite{qi08}.
The Dirac index is related to the $\eta$ invariant introduced by
Atiyah, Patodi and Singer\cite{ati73,ati75,ati75b,ati76}.
The $\eta$ invariant has also relation with the fermion number
that can be fractional in a soliton-Dirac fermion system.
New low-lying excitation modes would appear when fermions interact
with soliton-like objects such as domain walls, vortices,
kinks and monopoles\cite{jac81,cal85,wei12,man04}.
There also exist zero-energy bosonic modes on solitons\cite{man04,raj82,yan18},
and thus both bosonic and fermionic zero-energy modes will emerge in
the presence of solitons.
These exotic quantum states carry fermionic quantum numbers that
can be fractional\cite{jac76,su79,su80,gol81}.
The existence of Majorana zero modes has also been examined
in doped topological materials\cite{hos11,jia19}.

We expect that the quantization depends on a topological structure.
In superconductors, the magnetic flux is quantized as integer times
the unit quantum flux $\phi_0$.
There are, however, exceptions when superconductors have multi
components or form some geometric structure.
A fractional-flux quantum vortex (FFQV) may exist in a 
multi-component or multi-layer superconductor.
In fact, an FFQV has been observed in Nb thin
film superconducting bilayers recently\cite{tan18}.
This may raise a question about quantization.

In this paper we investigate zero-energy modes in a
vortex-fermion system and a fractional vortex-fermion system.  
The zero-energy mode is a Majorana
fermion mode in a Dirac semi-metal with vanishing chemical
potential.  The inclusion of non-zero chemical potential would
change the nature of excitation modes.
If a bilayer system including superconductors and a topological
insulator is synthesized, the Dirac fermion on the surface of the
topological insulator may cause a zero-energy mode in a vortex.
There are several superconductors that are 
suggested to be a topological superconductor of Dirac 
electrons\cite{kon19,iyo16,liu19}.
They are, for example, FeTe$_{1-x}$Se$_x$\cite{kon19},
CaKFe$_4$As$_4$\cite{iyo16,liu19}.
In (Bi$_{1-x}$Sb$_x$)$_2$Te$_3$ a surface Dirac electronic
state is suggested to be realized.
A vortex-Dirac fermion system may be formulated on a surface
of a junction of superconductors and a topological insulator.
Our discussion will include the case where there is a half-flux
quantum vortex (HFQV) that is associated with a kink in a
bilayer superconductor in a magnetic field.
A normalizable single-valued or two-valued fermion
zero-energy mode does not exist in the core of HFQV.

The index has been defined for Dirac operators.
The index of a Dirac operator is closely related to the
fermion number and $\eta$ invariant.  The fermion number
can be fractional in a Dirac system.
The Dirac index will have an additional contribution if
a scalar field has a singularity like a vortex.
The paper is organized as follows.  In Section 2 we examine
fermion zero-energy modes in a vortex-Dirac fermion system.
We show that we can identify the fermion zero-energy mode
as a Majorana mode when the chemical potential $\mu=0$.
In Section 3 we discuss the index of a Dirac operator and
fractional fermion number in a vortex-Dirac fermion system.
We give a summary in the lase section.

\section{Fermion zero-energy modes and solitons}

\subsection{A vortex-Dirac fermion model}

When Dirac fermions couple to a soliton, there may appear
localized fermion zero modes in a soliton.
Let us consider Dirac fermions in (1+2) dimensions where
Dirac fermions interact with a scalar field.
The Lagrangian is given by\cite{jac81}
\begin{equation}
\mathcal{L}= -\frac{1}{4}F_{\mu\nu}F^{\mu\nu}
+ \bar{\psi}\gamma^{\mu}(i\partial_{\mu}-qA_{\mu})\psi
-\frac{1}{2}ig\phi\bar{\psi}\psi^c+\frac{1}{2}ig^*\phi^*\bar{\psi^c}\psi,
\end{equation}
where $\psi$ is a two-component spinor and $q$ is the coupling
to the gauge field.  We use the notation $\bar{\psi}=\psi^{\dag}\gamma^0$.  
Usually we choose $q=e$ or $q=2e$ where $e$ is the electron charge.
We will choose $q=2e$ so that the index of the Dirac operator becomes
an integer since the Dirac index is the difference of the dimensions of
vector spaces.  This will be described in Section 3. 
This is related to the property that the magnetic flux is quantized as an 
integer times the quantum
unit $\phi_0=h/2|e|=\pi\hbar/|e|$.
$A_{\mu}$ is the abelian gauge field and $F_{\mu\nu}$ is the
field strength given by $F_{\mu\nu}=\partial_{\mu}A_{\nu}-\partial_{\nu}A_{\mu}$.
$\psi^c$ is the charge conjugate spinor given as
$\psi^c= C\bar{\psi}^T$ where $C$ is the charge conjugation matrix
and $T$ indicates the transposition.
$g$ is the coupling constant.
Dirac matrices are chosen as
\begin{equation}
\gamma^0= \sigma_3,~~\gamma^1= i\sigma_2,~~ \gamma^2= -i\sigma_1,
\end{equation}
and
\begin{equation}
C = i\gamma^0\gamma^2 = i\sigma_2.
\end{equation}
We use the Minkowski metric $(\eta^{\mu\nu})={\rm diag}(1,-1,-1)$.
For the representation
\begin{eqnarray}
\psi= \left(
\begin{array}{c}
\psi_1 \\
\psi_2 \\
\end{array}
\right),
\end{eqnarray}
the interaction term is written as
\begin{equation}
\mathcal{L}_{int}\equiv -\frac{i}{2}g\phi \bar{\psi}\psi^c
+\frac{i}{2}g\phi^*\bar{\psi^c}\psi
= ig\phi\psi_1^*\psi_2^*-ig\phi^*\psi_2\psi_1.
\end{equation}
$\mathcal{L}_{int}$ indicates the pairing interaction between
$\psi_1$ and $\psi_2$.  Thus $\mathcal{L}$ in eq.(1) represents
a superconductor model in (1+2) dimensions.

We assume that $A^0=0$ and
\begin{equation}
A^i(x,y)= \epsilon^{ij}\hat{{\bf r}}_j\frac{1}{2e}a(r),
\end{equation}
where ${\hat{\bf r}}={\bf r}/|{\bf r}|$ with ${\bf r}=(x,y)$ and
$r=|{\bf r}|$.  $a(r)$ is a function of the radial variable $r$.
The scalar field $\phi$ corresponds to the gap function and we
assume the form with the vorticity $Q$:
\begin{equation}
\phi({\bf r})= e^{iQ\theta}f(r),
\end{equation}
where $\theta$ is the angle variable $\theta=\tan^{-1}(y/x)$ and
$f(r)$ is a function of $r$.
In the conventional case $Q$ takes an integer value.  In this paper
we also consider the case where  $Q$ could take a non-integer 
value\cite{yan19a}.
We assume the asymptotic behaviors for $f(r)$ and $a(r)$ as follows:
\begin{eqnarray}
f(r) &\rightarrow& f_{\infty}~~~(r\rightarrow \infty)\\
     &\rightarrow& f_0r^{|Q|}~~~(r\rightarrow 0)\\
a(r) &\rightarrow& -Q/r~~~(r\rightarrow \infty)\\
     &\rightarrow& 0~~~(r\rightarrow 0).
\end{eqnarray}
Here $f_{\infty}$ and $f_0$ are constants.
We assume that $gf(r)\ge 0$.
Then the magnetic flux is given by
\begin{equation}
\Phi= -\int d^2xF_{12}= \int dxdyF_{xy}= \frac{\pi}{e}Q,
\end{equation}
for $F_{xy}=\partial_xA_y-\partial_yA_x$ where we set $A_x=A^1$ and $A_y=A^2$.
We use the unit $\hbar=c=1$ in this paper.

Let us consider fermion zero-energy modes in this system.
The equation of motion for $\psi$ is given by
\begin{equation}
i\partial_t\psi = \sigma_j\left( -i\partial_j-qA^j\right)
-g\phi\sigma_2\psi^*.
\end{equation}
The equation for the zero-energy mode is written as
\begin{equation}
\sigma_j(-i\partial_j-  A^j)\psi-g\phi\sigma_2\psi^* = 0.
\end{equation}
We set $D_j= \partial_j-ieA^j$ to obtain
\begin{align}
D_1+iD_2 &= e^{i\theta}\left( \partial_r+i\frac{1}{r}\partial_{\theta}
-a(r)\right),\\
D_1-iD_2 &= e^{-i\theta}\left( \partial_r-i\frac{1}{r}\partial_{\theta}
-a(r)\right),
\end{align}
for $x=r\cos\theta$ and $y=r\sin\theta$.
A solution $\psi$ is written in the form
\begin{eqnarray}
\psi= \left(
\begin{array}{c}
e^B\chi_1 \\
e^{-B}\chi_2 \\
\end{array}
\right),
\end{eqnarray}
where
\begin{equation}
B= \int_0^rdr'a(r').
\end{equation}
$\chi_1$ and $\chi_2$ should satisfy
\begin{align}
e^{i\theta}\left( \partial_r+\frac{i}{r}\partial_{\theta}\right)\chi_1
+gfe^{iQ\theta}\chi_1^*&= 0, \\
e^{-i\theta}\left( \partial_r+\frac{i}{r}\partial_{\theta}\right)\chi_2
-gfe^{iQ\theta}\chi_2^*&= 0. 
\end{align}

When $Q$ is an integer, there are $|Q|$ normalizable solutions\cite{jac81}.
This is easily shown by using the following Fourier decomposition:
\begin{align}
\chi_1 &= e^{i(Q-1)\theta/2}\sum_{\ell}e^{i\ell\theta}\chi_{1\ell},\\
\chi_2 &= e^{i(Q+1)\theta/2}\sum_{\ell}e^{i\ell\theta}\chi_{2\ell}.
\end{align}
We adopt that $\chi_{1\ell}$ and $\chi_{2\ell}$ are real.
Then we have the equations for $\chi_{1\ell}$ as
\begin{align}
\left( \partial_r-\frac{(Q-1)/2+\ell}{r}\right)\chi_{1\ell}
+gf\chi_{1,-\ell}&= 0,\\
\left( \partial_r-\frac{(Q-1)/2-\ell}{r}\right)\chi_{1,-\ell}
+gf\chi_{1\ell}&= 0.
\end{align}
Similarly the equations for $\chi_{2\ell}$ are
\begin{align}
\left( \partial_r+\frac{(Q+1)/2+\ell}{r}\right)\chi_{2\ell}
+gf\chi_{2,-\ell}&= 0,\\
\left( \partial_r+\frac{(Q+1)/2-\ell}{r}\right)\chi_{2,-\ell}
+gf\chi_{2\ell}&= 0.
\end{align}
The following conditions should be satisfied so that
$\chi_{1\ell}$ and $\chi_{1,-\ell}$ are regular at the origin:
\begin{equation}
-(Q-1)/2\leq\ell \leq (Q-1)/2.
\label{leq1}
\end{equation}
This indicates that $Q\ge 1$ and the allowed values of $\ell$ are as follows.
For $Q=1$, we have $\ell=0$.  For $Q=2$, $\ell=\pm 1/2$.
For $Q=3$, $\ell$ takes $-1$, 0 and 1, and so on.  This is shown in Table 1.
Hence there are $Q$ solutions for $\chi_1$.
The condition for $\chi_2$ reads
$(Q+1)/2\leq \ell\leq -(Q+1)/2$.  Thus $Q$ should be negative
and $\ell$ is in the range
\begin{equation}
-(|Q|-1)/2 \leq\ell \leq (|Q|-1)/2,~~~~Q\leq -1.
\label{leq2}
\end{equation}
Therefore $\chi_2$ vanishes when $\ell$ is in the range of eq.(\ref{leq1}),
and instead $\chi_1$ vanishes when $Q$ and $\ell$ satisfy eq.(\ref{leq2}).

\begin{table}
\caption{Allowed values of $\ell$ for positive integers $Q$.
$\ell$ takes half-integers when $Q$ is an even integer.
$m$ indicates a power of $\chi_{1\ell}$ for small $r\sim 0$.
}
\centering
\scalebox{.85}[.85]{
\begin{tabular}{ccc}
\hline
 \boldmath{$Q$}  & \boldmath{$\ell$} & \boldmath{$m\equiv (Q-1)/2+\ell$}   \\
\hline
1      &  0       &  0  \\ 
2      &  $-1/2$, $1/2$  &  0, 1   \\
3      &  $-1$, 0, 1  &  0, 1, 2 \\
4      &  $-3/2$, $-1/2$, 1/2, 3/2  &  0, 1, 2, 3 \\
5      &  $-2$, $-1$, 0, 1, 2  &  0, 1, 2, 3, 4 \\
\hline
\end{tabular}
}
\label{tab1}
\end{table}

When $\ell$ is non-zero, a pair of $\chi_{1\ell}$ and $\chi_{1,-\ell}$
or $\chi_{2\ell}$ and $\chi_{2,-\ell}$ contribute to a gapless
mode.
When $f$ vanishes, $\chi_{1\ell}$ is given by 
$\chi_{1\ell}\simeq r^{(Q-1)/2+\ell}$.
$\chi_{1\ell}$ satisfies the second-order differential equation:
\begin{eqnarray}
& \partial_r^2\chi_{1\ell}+\frac{1-Q}{r}\partial_r\chi_{1\ell}
\left( \frac{Q^2}{4}-\left(\ell-\frac{1}{2}\right)^2\right)
\frac{1}{r^2}\chi_{1\ell} \nonumber\\
& ~~~~~ +gf'(r)\chi_{1,-\ell}
-(gf)^2\chi_{1\ell} =0.
\end{eqnarray}
This is the second-order differential equation with a regular 
singular point\cite{pon62,cod84} if $f(r)$ is a regular function.
When $r$ is large, we neglect $1/r$ term in the equation to have
\begin{equation}
\chi_{1\ell}\simeq \chi_{1.-\ell} \simeq \exp\left(-\int_0^r gf(r')dr'\right).
\end{equation}
For small $r$, since $f(r)\rightarrow 0$ as $r\rightarrow 0$,
the behavior of solutions is determined by the indicial equation
given by
\begin{equation}
k^2-Qk+\left(\frac{Q}{2}\right)^2-\left(\ell-\frac{1}{2}\right)^2=0.
\end{equation}
There are two solutions for this equation:
\begin{equation}
k_1=\frac{Q-1}{2}+\ell,~~~~k_2=\frac{Q+1}{2}-\ell.
\end{equation}
Since $k_1-k_2\geq 0$ if and only if $\ell\geq 1/2$,
$\chi_{1\ell}$ exhibits the power behavior
\begin{equation}
\chi_{1\ell} \simeq r^{\frac{Q-1}{2}+\ell}\varphi(r),
\end{equation}
for $\ell\geq 1/2$, where $\varphi(r)$ is a non-singular function.
The power $(Q-1)/2+\ell$ coincides with that derived from eq.(23)
in the limit $f(r)\rightarrow  0$.

Let us examine the relation between the spinor $\psi$ and the Majorana
spinor. 
The zero-energy mode with $\ell=0$ for a positive odd integer $Q$ is given by
\begin{equation}
\chi_{1\ell=0} = r^{(Q-1)/2} \exp\left( -\int_0^r gf(\rho)d\rho \right),
\end{equation}
and $\chi_2=0$.  From this solution the Majorana fermion is
formulated as
\begin{eqnarray}
\psi_M = \psi+\psi^c =\left(
\begin{array}{c}
\xi \\
-\xi^* \\
\end{array}
\right),
\end{eqnarray}
where
\begin{equation}
\xi = e^B\chi_{1\ell=0}.
\end{equation}
Thus the fermion zero-energy mode can be regarded as the
Majorana mode.
The same argument applies for the zero-energy modes with $\ell\neq 0$.
The Majorana spinor is also made from $\psi$ for $\ell\neq 0$ since
$\chi_{2\ell}$ vanishes for $Q>0$.
Thus there can be $|Q|$ Majorana modes in general.

When $Q$ is a half-integer, $m\equiv (Q-1)/2+\ell$ must be also 
a half-integer so that
the wave function is a single-valued or two-valued function.  
For $Q=1/2$ no value of $\ell$ is allowed. 
Thus there is no normalizable and two-valued solution of the zero-energy
modes for $Q=1/2$.  For $Q=3/2$ we have $\ell=\pm 1/4$ or $m=0,1/2$.
It appears that there are $[Q]$ single-valued solutions for positive $Q$ where
$[Q]$ indicates the integer part of $Q$ (Gauss symbol).
For  half-integer $Q$,
we must have
\begin{equation}
[Q]>m> -\frac{1}{2},~~~ 2m\in {\bf Z},~~2Q\in {\bf Z}.
\end{equation}
For negative vorticity $Q<0$, we replace $Q$ by $|Q|$.
In fact, in the case of half-flux vortex with $Q=1/2$
we have a solution
\begin{equation}
\chi_1 = h(r)e^{-i\theta/4},
\end{equation}
and $\chi_2=0$.
For this ansatz we obtain
\begin{equation}
h(r) = r^{-\frac{1}{4}}\exp\left( -\int_0^rdr'gf(r')\right).
\end{equation}
This solution has a singularity  at $r\sim 0$ but can be normalized.
This solution, however, is not accepted because $\chi_1$ is not
a single-valued function.  In the system with a half-flux
quantum vortex, a wave function should be a single-valued or
two-valued function\cite{yan19}.
We show allowed values of $\ell$ and $m$ in Table 2.
There are $2[Q]=2Q-1$ solutions for $Q>0$ when including two-valued
solutions.

\begin{table}
\caption{Allowed values of $\ell$ for positive odd half-integers $Q$.
$2\ell$ takes half-integers in this case.
}
\centering
\scalebox{.85}[.85]{
\begin{tabular}{ccc}
\hline
 \boldmath{$Q$}  & \boldmath{$\ell$} & \boldmath{$m\equiv (Q-1)/2+\ell$}   \\
\hline
$1/2$   &  No solutions       &  No solutions  \\ 
$3/2$   &  $-\frac{1}{4}$, $\frac{1}{4}$  &  0, $\frac{1}{2}$   \\
$5/2$   &  $-\frac{3}{4}$, $-\frac{1}{4}$, $\frac{1}{4}$, $\frac{3}{4}$ 
 &  0, $\frac{1}{2}$, 1, $\frac{3}{2}$ \\
$7/2$   &  $-\frac{5}{4}$, $-\frac{3}{4}$, $-\frac{1}{4}$, 
$\frac{1}{4}$, $\frac{3}{4}$, $\frac{5}{4}$ &  0, $\frac{1}{2}$, 1, 
$\frac{3}{2}$, 2, $\frac{5}{2}$ \\
\hline
\end{tabular}
}
\label{tab2}
\end{table}


\subsection{Effect of the chemical potential}

We have discussed a Dirac semi-metal with vanishing chemical
potential $\mu=0$ so far.  In this subsection we examine a
Dirac metal by introducing the chemical potential.  
The wave function is a sum of the positive
and negative frequency parts:
\begin{equation}
\psi= e^{-iEt/\hbar}\psi_{+}({\bf r})
+e^{iEt/\hbar}\psi_{-}({\bf r}).
\end{equation}
The eigen-equation reads
\begin{align}
[\sigma_j(p_j-qA^j)-\mu]\psi_{+}-g\phi\sigma_2\psi_{-}^*&=E\psi_{+}, \\
[\sigma_j(p_j-qA^j)-\mu]\psi_{-}-g\phi\sigma_2\psi_{+}^*&= -E\psi_{-}.\\
\end{align}
We put
\begin{eqnarray}
\psi_{+}= \left(
\begin{array}{c}
e^B\psi_1 \\
e^{-B}i\psi_2 \\
\end{array}
\right), ~~~~
\psi_{-}= \left(
\begin{array}{c}
e^B\chi_1 \\
e^{-B}i\chi_2 \\
\end{array}
\right).
\end{eqnarray}
We neglect the magnetic field by assuming that the
Ginzburg-Landau parameter $\kappa$ is large, the equations
for $\psi_{+}$ and $\psi_{-}$ are represented as
\begin{align}
e^{i\theta}\left( \partial_r+\frac{i}{r}\partial_{\theta}\right)\psi_1
+g\phi\chi_1^* &= -(E+\mu)\psi_2, \\
e^{-i\theta}\left( \partial_r-\frac{i}{r}\partial_{\theta}\right)\psi_2
+g\phi\chi_2^* &= (E+\mu)\psi_1, \\
e^{i\theta}\left( \partial_r+\frac{i}{r}\partial_{\theta}\right)\chi_1
+g\phi\psi_1^* &= (E-\mu)\chi_2, \\
e^{-i\theta}\left( \partial_r-\frac{i}{r}\partial_{\theta}\right)\chi_2
+g\phi\psi_2^* &=-(E-\mu)\chi_1.
\end{align}
This set of equations is formally equivalent to the
Bogoliubov-de Gennes equation used for superconducting
graphene with two valleys\cite{bee06,jac08,kha09}.

We examine the zero-eigenvalue solution.
For $E=0$, we have a solution with $\chi_1=\psi_1$ and
$\chi_2=\psi_2$.  Then the equations read
\begin{align}
e^{i\theta}\left( \partial_r+\frac{i}{r}\partial_{\theta}\right)\psi_1
+g\phi\psi_1^* &= -\mu\psi_2, \\
e^{-i\theta}\left( \partial_r-\frac{i}{r}\partial_{\theta}\right)\psi_2
+g\phi\psi_2^* &= \mu\psi_1. 
\end{align}
We use the representation
\begin{align}
\psi_1 &= e^{i(Q-1)\theta/2}\sum_{\ell}e^{i\ell\theta}\psi_{1\ell},\\
\psi_2 &= e^{i(Q+1)\theta/2}\sum_{\ell}e^{i\ell\theta}\psi_{2\ell}.
\end{align}
The equations are given as
\begin{align}
\left( \partial_r-\frac{(Q-1)/2+\ell}{r}\right)\psi_{1\ell}
+gf\psi_{1,-\ell}&= -\mu\psi_{2\ell},\\
\left( \partial_r+\frac{(Q+1)/2+\ell}{r}\right)\psi_{2\ell}
+gf\psi_{2,-\ell}&= \mu\psi_{1\ell}.
\end{align}

In the limit $f\rightarrow 0$, $\psi_{1\ell}$ and $\psi_{2\ell}$
are given by Bessel functions:
\begin{align}
\psi_{1\ell}(r)\big|_{f\rightarrow 0} &= J_{\frac{Q-1}{2}+\ell}(|\mu| r), \\
\psi_{2\ell}(r)\big|_{f\rightarrow 0} &= J_{\frac{Q+1}{2}+\ell}(|\mu| r). 
\end{align}
For $\ell=0$, the solution in the presence of $gf$ is easily obtained as
\begin{align}
\psi_{1\ell=0}(r) &= \exp\left( -\int_0^r gf(r')dr'\right)
J_{\frac{Q-1}{2}}(|\mu| r), \\ 
\psi_{2\ell=0}(r) &= \exp\left( -\int_0^r gf(r')dr'\right)
J_{\frac{Q+1}{2}}(|\mu| r). 
\end{align}

In the limit $r\rightarrow 0$, since $f(r)\rightarrow 0$,
$\psi_{1\ell}$ and $\psi_{2\ell}$ approach Bessel functions shown above.
For large $r$, $r\rightarrow \infty$, we may neglect $1/r$ terms
so that we have
\begin{align}
\partial_r\psi_{1\ell}+gf\psi_{1,-\ell} & \simeq -\mu\psi_{2\ell},\\
\partial_r\psi_{2\ell}+gf\psi_{2,-\ell} & \simeq \mu\psi_{1\ell}.
\end{align}
Since the equations for $\psi_{j\ell}$ ($j=1,2$) are independent of
$\ell$, we assume that $\psi_{j\ell}=\psi_{j,-\ell}$.  Then
the asymptotic behaviors for large $r$ are
\begin{align}
\psi_{1\ell}& \simeq \cos(\mu r)\exp\left( -\int_0^r gf(r')dr'\right),\\
\psi_{2\ell}& \simeq \sin(\mu r)\exp\left( -\int_0^r gf(r')dr'\right),
\end{align}
or we have
\begin{align}
\psi_{1\ell}& \simeq \sin(\mu r)\exp\left( -\int_0^r gf(r')dr'\right),\\
\psi_{2\ell}& \simeq -\cos(\mu r)\exp\left( -\int_0^r gf(r')dr'\right).
\end{align}

\subsection{Dirac fermions and soliton fields}

Let us consider a model of Dirac fermions that couple with
scalar fields.  If scalar fields have a soliton-like structure,
a zero-energy mode would exist.
We consider the following Lagrangian
\begin{equation}
\mathcal{L}= -\frac{1}{4}F^{\mu\nu}F_{\mu\nu}
+ \bar{\psi}\gamma^{\mu}(i\partial_{\mu}-qA_{\mu})\psi
-g\bar{\psi}(\sigma_2\phi_1+\sigma_1\phi_2)\psi,
\end{equation}
where $\phi_1$ and $\phi_2$ are real scalar fields. 
The interaction term is written as
\begin{equation}
\mathcal{L}_{int}= ig\bar{\psi}\sigma_3M\psi,
\end{equation}
with
\begin{eqnarray}
M= \left(
\begin{array}{cc}
0 & \phi \\
\phi^* & 0 \\
\end{array}
\right),
\end{eqnarray}
where $\phi=\phi_1+i\phi_2$.

The equation for the zero-energy modes is
\begin{equation}
[\sigma_1(-i\partial_1-eA^1)+\sigma_2(-i\partial_2-eA^2)]\psi+gM\psi=0.
\end{equation}
We set the Fermi velocity $v_F=1$ for simplicity.
In a similar way, the wave function is written in the form
\begin{eqnarray}
\psi= \left(
\begin{array}{c}
e^B\chi_1 \\
e^{-B}\chi_2 \\
\end{array}
\right),
\end{eqnarray}
where
\begin{equation}
B= \int_0^r a(r')dr'.
\end{equation}
The equation for $(\chi_1,\chi_2)$ reads
\begin{align}
e^{i\theta}\left(\partial_r+\frac{i}{r}\partial_{\theta}\right)\chi_1
+g\phi^*\chi_1 &= 0, \\
e^{-i\theta}\left(\partial_r-\frac{i}{r}\partial_{\theta}\right)\chi_2
+g\phi\chi_2 &= 0. 
\end{align}
The gap function is parametrized as
\begin{equation}
\phi({\bf r})= e^{-in\theta}|\phi(r)|\equiv e^{-in\theta}f(r).
\end{equation}
We assume that $gf(r)>0$.
\begin{align}
\chi_1({\bf r}) &= \sum_{\ell\in {\bf Z}}e^{i\ell\theta}u_{\ell}(r), \\
\chi_2({\bf r}) &= \sum_{\ell\in {\bf Z}}e^{i\ell\theta}w_{\ell}(r),
\end{align}
where $\ell$ takes all the integer values.
We set $w_{\ell}=iv_{\ell}$, and then
the equations for fermion zero-energy modes with $E=0$ read
\begin{align}
\left( \partial_r-\frac{\ell}{r}\right)u_{\ell}(r)
+gf(r)u_{\ell-n+1}(r) &= 0, \\
\left( \partial_r+\frac{\ell}{r}\right)v_{\ell}(r)
+gf(r)v_{\ell+n-1}(r) &= 0.
\end{align}
For the vorticity $n=1$, we have
\begin{align}
\left( \partial_r-\frac{\ell}{r}\right)u_{\ell}(r)
+gf(r)u_{\ell}(r) &= 0, \\
\left( \partial_r+\frac{\ell}{r}\right)v_{\ell}(r)
+gf(r)v_{\ell}(r) &= 0.
\end{align}
The solutions are written as
\begin{align}
u_{\ell}&= a_{\ell}r^{\ell}\exp\left( -\int_0^rgf(r')dr'\right),\\
v_{\ell}&= b_{\ell}r^{-\ell}\exp\left( -\int_0^r gf(r')dr' \right),
\end{align}
where $a_{\ell}$ and $b_{\ell}$ are normalization constants.
For $\ell=0$ 
\begin{equation}
u_0(r)=v_0(r)= \exp\left( -\int_0^r gf(r')dr' \right).
\end{equation}
We must have $u_{-\ell}= v_{\ell}$ when $\ell$ is replaced by $-\ell$.
The normalizable wave function that is regular at the origin is 
written as
\begin{align}
\chi_1 &= \sum_{\ell\ge 0}e^{i\ell\theta}a_{\ell}r^{\ell}
\exp\left( -\int_0^r gf(r')dr' \right), \\
\chi_2 &= i\sum_{\ell\le 0}e^{i\ell\theta}a_{-\ell}r^{-\ell}
\exp\left( -\int_0^r gf(r')dr' \right). 
\end{align}
This indicates that $\chi_2/i$ is the complex conjugate of $\chi_1$:
\begin{equation}
\chi_2 = i\chi_1^*.
\end{equation}
When we neglect the magnetic field,
$\psi$ is given as
\begin{eqnarray}
\psi = \left(
\begin{array}{c}
\chi_1 \\
i\chi_1^* \\
\end{array}
\right).
\end{eqnarray}
By multiplying $\psi$ by a phase factor $e^{i\pi/4}$, $\psi$ is
written in the form
\begin{eqnarray}
\psi = \left(
\begin{array}{c}
e^{i\pi/4}\chi_1 \\
-e^{-i\pi/4}\chi_1^* \\
\end{array}
\right)\equiv \left(
\begin{array}{c}
\xi \\
-\xi^* \\
\end{array}
\right),
\end{eqnarray}
where we set $\xi= e^{i\pi/4}\chi_1$.
Hence we have obtained the Majorana spinor satisfying
\begin{equation}
\psi = \psi^c.
\end{equation}
We reached the conclusion that the fermion zero-energy mode is
represented by the Majorana spinor.

\section{Dirac operator and fractional fermion number}

\subsection{Index of the Dirac operator}

Let us consider the Dirac Hamiltonian given as
\begin{equation}
H= \sigma_j(-i\partial_j-qA^j)+gM+\sigma_3m,
\end{equation}
where $M$ is the matrix of the gap function in eq.(67) and
the mass $m$ is a constant.  $H$ is written as
\begin{eqnarray}
H= \left(
\begin{array}{cc}
m & D+g\Delta \\
D^{\dag}+g\Delta^* & -m \\
\end{array}
\right),
\end{eqnarray}
where
\begin{equation}
D= -i\frac{\partial}{\partial x}-qA_x-\frac{\partial}{\partial y}
+iqA_y.
\end{equation}
We put
\begin{eqnarray}
\cancel D_{\Delta} = \left(
\begin{array}{cc}
0  &  D+g\Delta \\
D^{\dag}+g\Delta^* &  0 \\
\end{array}
\right).
\end{eqnarray}
Since $\cancel D_{\Delta}$ anticommutes with $\sigma_3$, we can
define the index by
\begin{equation}
{\rm Ind}(\cancel D_{\Delta}) := {\rm Tr}_{\cancel D_{\Delta}\psi=0}\sigma_3.
\end{equation} 
Here, the trace Tr is evaluated in the space 
Ker$\cancel D_{\Delta}=\{\psi| \cancel D_{\Delta}\psi=0\}$.
This definition means
\begin{equation}
{\rm Ind}(\cancel D_{\Delta})= {\rm dimKer}D_{\Delta}^{\dag}
-{\rm dimKer}D_{\Delta},
\end{equation}
where $D_{\Delta}=D+g\Delta$ and $D_{\Delta}^{\dag}=D^{\dag}+g\Delta^*$.
The index is represented as by introducing the cutoff:
\begin{equation}
{\rm Ind}(\cancel D_{\Delta})= \lim_{\Lambda\rightarrow\infty}{\rm Tr}
\sigma_3e^{-\cancel D_{\Delta}^2/\Lambda^2}.
\end{equation}
Then Ind$(\cancel D_{\Delta})$ is calculated as
\begin{align}
{\rm Ind}(\cancel D_{\Delta}) &= \lim_{\Lambda\rightarrow\infty}
\int\frac{d^dk}{(2\pi)^d}{\rm tr}\langle k|\sigma_3
e^{-\cancel D_{\Delta}^2/\Lambda^2}|k\rangle \nonumber\\
&= \lim_{\Lambda\rightarrow\infty}\int\frac{d^dk}{(2\pi)^d}
\int d^dx {\rm tr}e^{-ik\cdot x}e^{-\cancel D_{\Delta}^2/\Lambda^2}
e^{ik\cdot x},
\end{align}.
where the tr indicates the trace operation with respect to $2\times 2$
matrices.
We use the formula
\begin{equation}
e^{-ik\cdot x}f(\partial_{\mu})e^{ik\cdot x}\psi=f(\partial_{\mu}+ik_{\mu})
\psi,
\end{equation}
for a function $f$, so that we have
\begin{equation}
{\rm Ind}(\cancel D_{\Delta}) = \lim_{\Lambda\rightarrow\infty}
\int\frac{d^dk}{(2\pi)^d}\int d^dx{\rm tr}\left(\sigma_3
e^{-\cancel D_{\Delta}^2/\Lambda^2}
\Big|_{\partial_{\mu}\rightarrow \partial_{\mu}+ik_{\mu}}\right).
\end{equation}
$\cancel D_{\Delta}^2$ is given as
\begin{eqnarray}
\cancel D_{\Delta}^2 = \left(
\begin{array}{cc}
(D+g\Delta)(D^{\dag}+g\Delta^*) & 0 \\
0 & (D^{\dag}+g\Delta^*)(D+g\Delta) \\
\end{array}
\right). \nonumber\\
\end{eqnarray}
The matrix elements are evaluated as
\begin{eqnarray}
& (D+g\Delta)(D^{\dag}+g\Delta^*)
\Big|_{\partial_{\mu}\rightarrow\partial_{\mu}+ik_{\mu}} \nonumber\\
&= k'^2_x+k'_x g\Delta_1+g\Delta_1 k'_x+\Delta_1^2
+ k'^2_y-k'_yg\Delta_2 \nonumber\\ 
&-g\Delta_2 k'_y+\Delta_2^2 
 -q[\partial_x,A_y]-q[A_x,\partial_y] \nonumber\\
&-i[k'_x,\Delta_2]
-i[k'_y,\Delta_1] \nonumber\\
&= (k'_x+\Delta_1)^2+(k'_y-\Delta_2)^2-q(\partial_xA_y-\partial_yA_x)
\nonumber\\
&-(\partial_x\Delta_2)-(\partial_y\Delta_1),
\end{eqnarray}
where we set $k'_j=k_j-i\partial_j-qA_j$ for $j=x$ and $y$.
In two-space dimensions $d=2$,  this results in
\begin{align}
{\rm Ind}(\cancel D_{\Delta}) &= \lim_{\Lambda\rightarrow\infty}
\frac{1}{4\pi}\int d^2x \Lambda^2{\rm tr}\sigma_3 e^{F/\Lambda^2}
\nonumber\\
&= \frac{1}{2\pi}\int d^2x \left( qF_{xy}+\partial_x\Delta_2
+\partial_y\Delta_1\right) \nonumber\\
&= {\rm Ind}(\cancel D)+{\rm Ind}(\Delta) \nonumber\\
&= \frac{e}{\pi}\Phi+{\rm Ind}(\Delta),
\end{align}
where $F_{xy}=\partial_x A_y-\partial_y A_x$ and
\begin{eqnarray}
F= \left(
\begin{array}{cc}
qF_{xy}+\partial_x\Delta_2+\partial_y\Delta_1 & 0 \\
0 & -qF_{xy}-\partial_x\Delta_2-\partial_y\Delta_1 \\ 
\end{array}
\right). \nonumber\\
\end{eqnarray}
We defined
\begin{equation}
{\rm Ind}(\Delta)= \frac{1}{2\pi}\int d^2x(\partial_x\Delta_2
+\partial_y\Delta_1)= -\frac{1}{2\pi}\int d^2x {\rm rot}\vec{\Delta},
\end{equation}
for $\vec{\Delta}=(\Delta_1,-\Delta_2)$.
This formula indicates that the Dirac index becomes non-zero if a
scalar field is singular even when no magnetic field is applied. 
When $\Delta=\Delta_1+i\Delta_2$ is not singular in two-space dimensions,
the integral concerning the gap functions vanishes.
In this case we have ${\rm Ind}(\cancel D_{\Delta})={\rm Ind}(\cancel D)$:
\begin{equation}
{\rm Ind}(\cancel D_{\Delta})= {\rm Ind}(\cancel D)
=\frac{q}{2\pi}= \frac{e}{\pi}\Phi.
\end{equation}
When the vorticity is $n=1$, $\Phi$ is given by the unit flux
$\Phi=\pi/e=-\phi_0$ where $\phi_0=\pi/|e|$ ($\hbar=1$).  This leads to
\begin{equation}
{\rm Ind}(\cancel D_{\Delta})= 1.
\end{equation}
Then we have
\begin{equation}
{\rm dimKer}D_{\Delta}^{\dag}-{\rm dimKer}D_{\Delta}=1.
\end{equation}
In fact, for positive angular momentum $\ell$, we have a zero-energy
normalizable solution $\psi$ satisfying $D_{\Delta}^{\dag}\psi=0$ for the 
Hamiltonian $H$ with $m=0$, and a solution for $D_{\Delta}\psi=0$
is not normalizable due to a singularity at the origin.
Since the solution of $D_{\Delta}^{\dag}\psi=0$ is also an eigenstate of
$\sigma_3$, this zero-mode can be regarded as a Majorana fermion.

\subsection{Fractional fermion number}

Let us consider the fermion number defined by
\begin{equation}
N= \int d^2x :\psi^{\dag}({\bf r})\psi({\bf r}):
= \frac{1}{2}\int d^2x [\psi^{\dag}({\bf r}),\psi({\bf r})],
\end{equation}
for ${\bf r}=(x,y)$ where $:\cdots :$ indicates the normal ordering.
$N$ is related to the eta invariant defined as
\begin{equation}
\eta_H(s)= \sum_{\lambda}{\rm sign}(\lambda)|\lambda|^{-s},
\end{equation}
where $\lambda$'s are eigenvalues of $H$.
The fermion number $N$ is given as\cite{nie85}
\begin{equation}
N= -\frac{1}{2}\eta_H(0).
\end{equation}
There is the relation between $\eta_{\cancel D}$ and 
Ind$(\cancel D)$\cite{nie85,nie86}:
\begin{equation}
\eta_{\cancel D}(0)= -{\rm Ind}(\cancel D).
\end{equation}
This is generalized to
\begin{equation}
\eta_{\cancel D_{\Delta}}(0)= -{\rm Ind}(\cancel D_{\Delta}).
\end{equation}
Then the fermion number in the massless limit is
\begin{equation}
N= \frac{q}{4\pi}\Phi = \frac{e}{2\pi}\Phi.
\end{equation}
When the flux $\Phi$ is $-n$ times the unit flux quantum,
we have the fractional fermion number
\begin{equation}
N= \frac{1}{2}n.
\end{equation}
When $m$ is finite, $N$ is given by
\begin{equation}
N= \frac{q}{4\pi}\frac{m}{|m|}\Phi
= -\frac{q}{8\pi}\frac{m}{|m|}\int d^2x \epsilon^{ij}F_{ij}.
\end{equation}
$N$ is written as
\begin{equation}
N= \int d^2x j^0.
\end{equation}
by introducing the fermion current $j^{\mu}$.
This suggests that the additional effective action is formulated as
\begin{equation}
\Delta S= {\rm sign}(m)\frac{q^2}{16\pi}\epsilon^{\mu\nu\sigma}
\int d^3x F_{\mu\nu}A_{\sigma},
\end{equation}
because of $\delta S/\delta A_{\mu}= -q\langle\bar{\psi}\gamma^{\mu}\psi\rangle
= -q\langle j^{\mu}\rangle$ for the action $S$.
Hence the Chern-Simons term is induced in a Dirac-vortex system.
This may be realized on the surface of a junction of a 
superconductor and a topological insulator.

\subsection{Fractional vortex and Dirac index}

Let us turn to the case of fractional-flux quantum vortex, that is,
the fractional vorticity $Q$, especially the case of half-flux
quantum vortex.  
The index ${\rm Ind}(\cancel D)$ equals $Q$
for $\Phi=-Q\phi_0$:
\begin{equation}
{\rm Ind}(\cancel D)= Q.
\end{equation}
${\rm Ind}(\cancel D)$ should be an integer since the index is
only the difference of dimensions of vector spaces.
${\rm Ind}(\cancel D_{\Delta})$ has a contribution from the
gap function because the phase of $\Delta$ has a singularity
on the kink\cite{yan12}.
The half-flux quantum vortex exists associated with the kink in
the phase space, where the kink is a one-dimensional object.
We here give a comment on the kink in a multiband superconductor.
The kink state may be unstable because of the energy cost when
the field changes rapidly.
In other words, the superconducting current flows between the
layer, which may cause a force to the magnetic flux vortex.
This may bring about a new effect on the zero modes in the vortex.
We, however, neglect this effect in this paper.

We adopt that the gap function is given as
\begin{equation}
\Delta({\bf r})= \Delta_0(r)e^{-i\phi(\theta)},
\end{equation}
where $\phi(\theta)$ has a step-function-like singularity,
\begin{equation}
\phi(\theta)= \frac{1}{2}\theta+\pi H(\theta),
\end{equation}
near the origin $-\pi<\theta<\pi$.
$H(\theta)$ indicates the Heaviside step function.
We assume $\Delta_0(r)=\Delta_{\infty}\tanh(r/\xi)$.
Then we calculate
\begin{eqnarray}
&\int dxdy\left( \partial_x\Delta_2+\partial_y\Delta_1\right)~~~ \nonumber\\ 
&=  -\int dxdy \Delta_0(r)\left( \cos\phi\cdot \phi'(\theta)\partial_x\theta
+ \sin\phi\cdot\phi'(\theta)\partial_y\theta \right) \nonumber\\
&= -\int_0^R dr\int_{-\pi}^{\pi}d\theta \Delta_0(r)\phi'(\theta)
\left( -\cos\phi\sin\theta+\sin\phi\cos\theta \right) \nonumber\\
&= -\pi\Delta_{\infty}\sin\phi(0)\xi\ln\cosh(R/\xi)). 
\end{eqnarray}
We take the cutoff $R$ so that $\ln\cosh(R/\xi)\simeq 1$ and
$\phi(0)=\pi/2$.
Since $\xi\simeq 1/\Delta_{\infty} (=\hbar v_F/\Delta_{\infty}$,
we have
\begin{equation}
\int dxdy\left( \partial_x\Delta_2+\partial_y\Delta_1\right) = -\pi.
\end{equation}
This indicates
\begin{equation}
{\rm Ind}(\cancel D_{\Delta})= {\rm Ind}(\cancel D)+{\rm Ind}(\Delta)
= Q-\frac{1}{2}.
\end{equation}
Thus ${\rm Ind}(\cancel D_{\Delta})$ becomes an integer
with the contribution from the kink for the half-flux vortex.

\subsection{Fermion number and kinks}

The existence of a fermion zero-energy mode is related to the
fractional fermion number.
Let us examine the (1+2)-dimensional model of Dirac fermions
that couples to a scalar field with kink structure.
The Lagrangian is given as
\begin{equation}
\mathcal{L}= -\frac{1}{4}F^{\mu\nu}F_{\mu\nu}
+ \bar{\psi}\gamma^{\mu}(i\partial_{\mu}-qA_{\mu})\psi
-\bar{\psi}(m+\gamma^1\phi_1)\psi,
\end{equation}
where $\phi_1$ is a real scalar field. 
We assume that $\phi_1$ represents a kink solution with the
asymptotic behavior,
\begin{eqnarray}
\phi_1(x)\rightarrow v~~ {\rm as}~x\rightarrow\infty,\\
\phi_1(x)\rightarrow -v~~ {\rm as}~x\rightarrow -\infty.
\end{eqnarray}
The kink is a one-dimensional object depending on one
variable and is situated outside the region where the vortex
exists.  Then the fermion number is a sum of two contributions
from vortex and kink:
\begin{equation}
N= {\rm Ind}(\cancel D)+N_{kink}.
\end{equation}
$N_{kink}$ is given by the Goldstone-Wilczek formula:
\begin{equation}
N_{kink}= -\frac{1}{2\pi}\left(
\tan^{-1}\left(\frac{\phi_1(\infty)}{m}\right)
-\tan^{-1}\left(\frac{\phi_1(-\infty)}{m} \right)\right).
\end{equation}
Then, in the limit $m\rightarrow 0$ for $v>0$, we have
\begin{equation}
N= {\rm sign}(m)\Big[ \frac{q}{4\pi}\Phi-\frac{1}{2}\Big]
=-{\rm sign}\left( \frac{1}{2}Q+\frac{1}{2}\right),
\end{equation}
where the flux is given as $\Phi=-Q\phi_0$.
For $v<0$,
\begin{equation}
N =-{\rm sign}\left( \frac{1}{2}Q-\frac{1}{2}\right),
\end{equation}

\section{Summary}

We have investigated fermion zero-energy modes and the
index of the Dirac operator in vortex-Dirac fermion systems
in (1+2) dimensions.
Dirac fermions play an important role in many electron
systems such as topological insulators, topological
superconductors, graphene\cite{mcc56,slo58,and05} and also Kondo 
systems\cite{yan12a,yan15,yan15b}.  
A vortex-Dirac fermion system may be
realized on the surface of a topological insulator in a
junction of superconductors and topological insulators.
We have shown that a fermion zero-energy mode exists
in a vortex-fermion system and in a soliton-fermion system.  The zero-energy
modes are described by Majorana fermions in a Dirac semi-metal ($\mu=0$).
The quasi-particle energy level $\epsilon_n=(n+1/2)\hbar\omega_0$
in the vortex core of conventional superconductors shifts to
$\epsilon_n=n\hbar\omega_0$ in Dirac superconductors.
We have also shown that there is no fermion zero mode in a vortex with
fractional vorticity less than unity since wave function
has a singularity at the origin or becomes a
multi-valued function.
There is a contribution to the index of a Dirac operator when
the scalar field has a soliton-like structure with singularity.
Lastly we give a comment that we neglected the non-equilibrium
dynamics that are caused by the superconducting current flow
between the layer brought about by the kink in a superconducting
bilayer.

This work was~supported by a Grant-in-Aid for Scientific
Research from the Ministry of Education, Culture, Sports, Science and
Technology of Japan (Grant No. 17K05559).
A part of the computations was supported by the Supercomputer
Center of the Institute for Solid State Physics, the~University of~Tokyo.


\end{document}